\documentclass[aps,preprint,superscriptaddress,journal=prl]{revtex4-2}
\usepackage{amsmath,amsthm,amssymb,amsfonts}
\usepackage{siunitx}
\usepackage{color}
\usepackage{hyperref}
\usepackage{diagbox}
\usepackage{graphicx} 
\usepackage{siunitx}
\usepackage{textcomp}
\usepackage{gensymb}
\usepackage{physics}
\hypersetup{colorlinks=true,linkcolor=blue,anchorcolor=blue,citecolor=blue}

\begin{document}

\title{\textbf{Revisiting Ferroelectricity Beyond Polar Space Groups}}

\author{Yudi Yang}
\thanks{These two authors contributed equally}
\affiliation{Department of Physics, School of Science, Westlake University, Hangzhou 310030, China}
\author{Changming Ke}
\thanks{These two authors contributed equally}
\affiliation{Department of Physics, School of Science, Westlake University, Hangzhou  310030, China}
\author{Shi Liu}
\email{liushi@westlake.edu.cn}
\affiliation{Department of Physics, School of Science, Westlake University, Hangzhou 310030, China}
\affiliation{Institute of Natural Sciences, Westlake Institute for Advanced Study, Hangzhou 310024, China}


\begin{abstract}
Ferroelectricity, a hallmark of spontaneous inversion-symmetry breaking, has been a central concept in condensed matter physics and functional materials research, yet recent discoveries are revealing that switchable polarization can emerge in forms far richer than allowed by the conventional symmetry-based paradigm.  Fractional quantum ferroelectricity and ionic-conductor ferroelectricity challenge the long-standing association of ferroelectricity exclusively with polar space groups. In this Review, we reconcile these emerging phenomena within the Berry-phase modern theory of polarization. We emphasize that polarization in insulating periodic crystals is not a single-valued vector, but a multivalued lattice quantity defined modulo a polarization quantum. Consequently, nonpolar crystals may possess nonzero formal polarization, and adiabatic paths connecting symmetry-equivalent structures can produce quantized changes in polarization without violating symmetry principles. The symmetry of this multivalued formal polarization is governed by a generalized Neumann principle. We further show that the large polarization changes induced by long-range ion migration in both fractional quantum ferroelectrics and ionic-conductor ferroelectrics can be naturally understood through the topological definition of oxidation state, which links ionic transport to quantized charge transfer and polarization change. We discuss the physical accessibility of these unconventional polarization states, highlighting the roles of switching pathways, boundary conditions, and domain-wall dynamics, particularly in systems such as $\alpha$-In$_2$Se$_3$. Finally, we suggest that the most promising functionality of these materials may lie not in conventional bulk ferroelectric switching, but in the creation and control of charged interfaces and domain walls arising from discontinuities in formal polarization.
\end{abstract}
\maketitle
\clearpage
\section{Introduction}
The origin of ferroelectricity can be traced to Joseph Valasek’s seminal 1921 discovery of dielectric hysteresis in Rochelle salt~\cite{Valasek21p475}, an observation he explicitly compared to the behavior of ferromagnets. This finding laid the foundation for the standard definition of ferroelectrics. As articulated in the influential book by Lines and Glass~\cite{Lines77}, a ferroelectric is a material with two or more distinct orientational states, each associated with a non-zero spontaneous polarization that can be reversed by an external electric field. Because polarization is traditionally defined as the dipole moment per unit volume, a vector quantity, it became widely accepted that spontaneous polarization must break spatial inversion symmetry, mirroring the role of time-reversal symmetry breaking in ferromagnets. As a result, the concept of ferroelectricity became closely linked to crystal symmetry, particularly through Neumann’s Principle~\cite{Neumann85book}, which states that a crystal can exhibit a spontaneous vector property only if it belongs to one of the ten polar point groups. This symmetry-based framework proved highly effective in describing ferroelectrics. It not only supported the development of Landau theory for the ferroelectric–paraelectric phase transition but also guided the identification and design of key ferroelectric materials~\cite{Ginzburg45p2949}, such as the perovskite barium titanate (BaTiO$_3$)~\cite{Junquera03p506, 
Cohen92p136,  
Saha00p8828}.
The relationship between ferroelectricity and symmetry was therefore viewed as both robust and comprehensive.

While the analogy between ferroelectricity and ferromagnetism has played a key role in understanding these materials, it obscures a key conceptual difference. Unlike the magnetic moment, an intrinsic and well-defined property of electron spin, electric polarization in a periodic crystal is not a uniquely defined local quantity. This limitation is particularly evident in light of recent proposals for unconventional ferroelectric states that appear to defy standard symmetry-based definitions. One such example is provided by fractional quantum ferroelectrics (FQFEs)~\cite{Ji24p1,Yu25p016801}, in which spontaneous polarization can arise even in crystals whose symmetries would forbid non-zero vector properties. Similarly, the concept of quantized ferroelectricity in ionic conductors, referred to here as ionic conductor ferroelectrics (ICFEs)~\cite{Wang24p3885}, proposes that long-range ionic displacements can generate quantized changes in polarization. Both concepts challenge the conventional view that switchable polarization must originate from one of the ten polar point groups. This, in turn, raises questions about the universality of Neumann's principle in describing all forms of switchable polarization.

In this Review, we examine these conceptual contradictions through the framework of the modern theory of polarization~\cite{KingSmith93p1651, Resta92p51, Resta94p899, Spaldin12p2}. Within this theory, polarization is not a classical vector but a lattice-valued, multivalued quantity defined modulo a quantum. Following the formalism developed by Resta and Vanderbilt, we draw a distinction between two key concepts: (1) \textit{formal polarization}, determined by the Berry phase of the electronic wavefunction and inherently multivalued, and (2) \textit{effective polarization}, defined as the change in polarization relative to a centrosymmetric reference state, which manifests physically as a measurable flow of charge.
When viewed through this lens, the unusual properties of FQFEs and ICFEs no longer appear paradoxical or in conflict with symmetry principles. Rather, they reflect a broader and more nuanced understanding of what it means for a material to exhibit switchable polarization. By distinguishing between formal and effective polarization, we return to the core of Valasek’s original insight: polarization is ultimately about the measurable movement of charge. To accurately classify these unconventional systems, we must move beyond static symmetry constraints and instead consider ion-transport dynamics, the influence of boundary conditions, and the functional consequences of polarization switching in real materials.

This Review is structured as follows. We begin by revisiting the modern theory of polarization (MTP), highlighting the multivalued and quantized nature of formal polarization, along with its topological character and symmetry constraints. We also clarify the relationship between formal and effective polarization, emphasizing how the latter relates to physically measurable quantities.
Section III applies this framework to FQFEs, discusses the generalized Neumann principle, and critically examines the physical mechanisms at play in materials like \(\alpha\)-In\(_2\)Se\(_3\), with a focus on domain-wall dynamics. In Section IV, we extend the analysis to ICFEs, linking long-range ionic transport to quantized polarization changes and exploring the practical challenges of device implementation. Finally, Section V synthesizes these concepts, proposing that the true utility of these materials lies in the engineering of charged interfaces and topological boundaries rather than in bulk switching, and offers a unified conceptual framework for what we term \textit{topological ionics}.

\section{Modern Theory of Polarization}
This section outlines the formal derivation of polarization in crystalline insulators and its link to the Berry phase~\cite{KingSmith93p1651, Resta94p899, Vanderbilt18}. Those already familiar with the MTP may wish to skip the detailed derivations in Section II.A and proceed directly to the later sections. 

\subsection{Formal Polarization}
In finite systems, macroscopic polarization ($\mathbf{P}$) is intuitively defined as the dipole moment ($\mathbf{d}$) per unit volume ($V$), $\mathbf{P} = \mathbf{d}/{V} = {V}^{-1} \int \mathbf{r} \rho(\mathbf{r}) \dd\mathbf{r}$, 
where $\mathbf{r}$ denotes the position vector and $\rho(\mathbf{r})$ is the total charge density. For a crystalline insulator under periodic boundary conditions, consider a supercell containing \(N\) primitive cells, each of volume \(\Omega\). The polarization is formally separated into ionic and electronic contributions:
\begin{equation}
\mathbf{P} = \mathbf{P}_\text{ion} + \mathbf{P}_\text{elec} = \frac{e}{\Omega} \sum_{s}^{\text{ions}} Z_s \mathbf{r}_s - \frac{e}{N\Omega} \sum_{\mathbf{k},n \in \text{occ}} \int \mathbf{r} |\psi_{\mathbf{k}n}(\mathbf{r})|^2 \dd\mathbf{r}
\label{eq:convP}
\end{equation}
Here, the ionic term treats the nuclei as classical point charges, with the sum taken over the ions in a chosen primitive cell. \(Z_s\) is the integer charge of the \(s\)-th ion, and \(\mathbf{r}_s\) is its position within the cell, measured relative to the chosen unit-cell origin. The electronic term is evaluated over the occupied Kohn--Sham (KS) orbitals \(\psi_{\mathbf{k}n}(\mathbf{r})\), where \(\mathbf{k}\) is the crystal momentum in the Brillouin zone and \(n\) labels the band.

\begin{figure}[h]
    \centering
    \includegraphics[width=\textwidth]{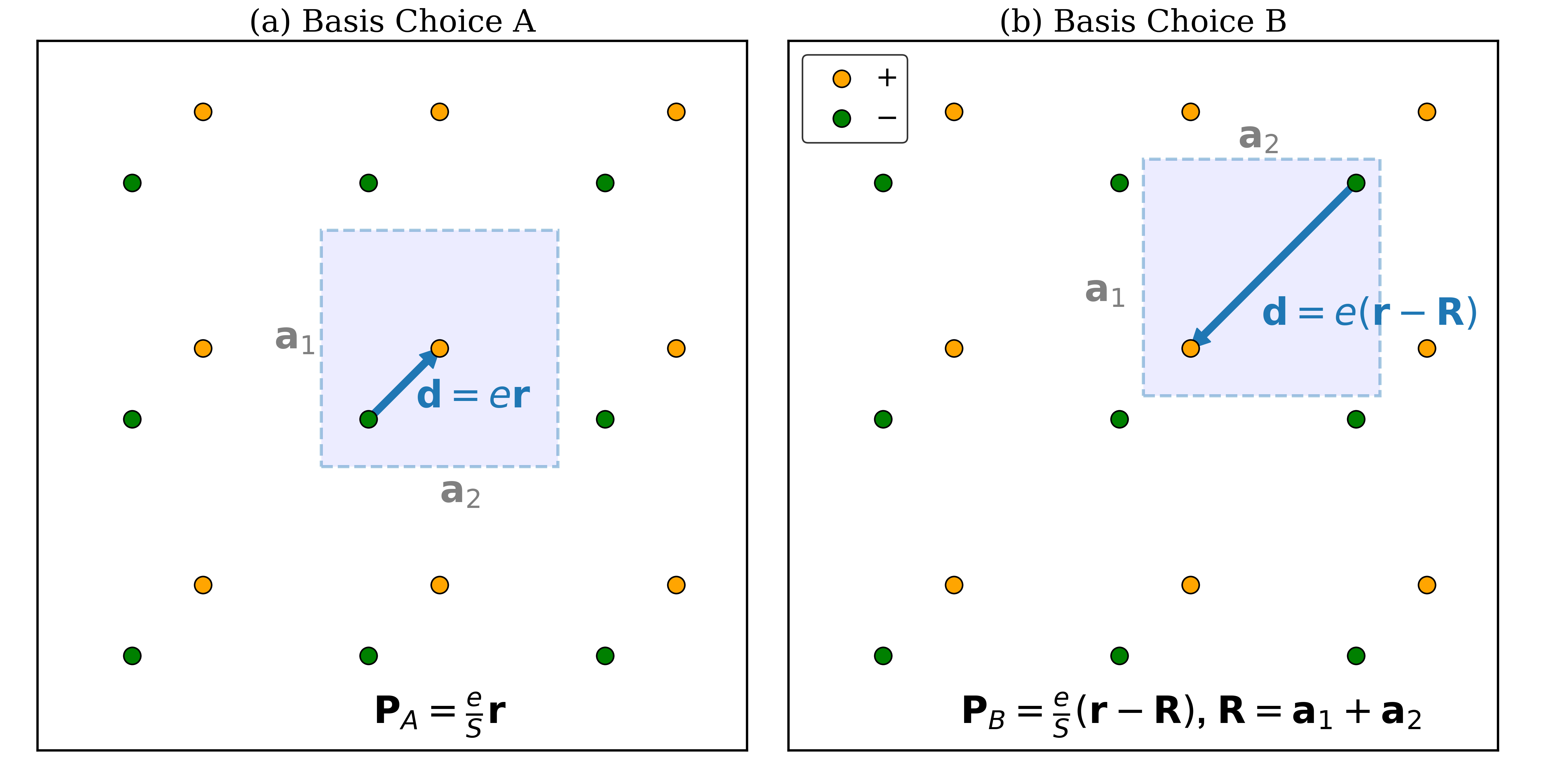}
    \caption{Multivalued formal polarization in a two-dimensional crystal of classical point charges. (a) Unit cell A: the unit cell is chosen so that the anion (green) is paired with the cation (orange) to its northeast. (b) Unit cell B: shifting the unit-cell boundary pairs the anion with the cation to its southwest. The resulting formal polarization changes by \(e\mathbf{R}/S\), where \(S\) is the unit-cell area.}
    \label{fig:classical_ambiguity}
\end{figure}

However, this conventional definition becomes problematic in periodic crystals. First, the choice of unit-cell boundary is arbitrary: shifting the boundary changes the net ionic dipole by a lattice vector, making \(\mathbf{P}_{\mathrm{ion}}\) ambiguous up to integer multiples of \(e\mathbf{R}/\Omega\), where \(\mathbf{R}\) is a lattice vector (see Fig.~\ref{fig:classical_ambiguity}). More fundamentally, in the thermodynamic limit (\(N \rightarrow \infty\)), the position operator \(\mathbf{r}\) is unbounded, so the electronic contribution cannot be defined in a mathematically meaningful way through a direct spatial integral. These difficulties motivate the development of the MTP.

The MTP treats polarization not as a static quantity determined directly from the bulk charge density, but as a \textit{dynamical} quantity defined through the time-integrated adiabatic current generated during a continuous insulating evolution of the system. This interpretation is closely aligned with experimental practice. In a typical ferroelectric measurement, the material is placed in a capacitor structure, with the ferroelectric film sandwiched between two metallic electrodes. When an external voltage is slowly varied, the polarization inside the ferroelectric changes, which modifies the bound charge at the film surfaces. To maintain electrostatic equilibrium, compensating screening charge is redistributed on the electrodes through the external circuit. The resulting current measured in the circuit therefore reflects the flow of screening charge induced by the change in polarization. By integrating this current over time, one obtains the change in polarization between two states. Repeating the measurement during field cycling yields the characteristic polarization--electric field hysteresis loop. Experimentally, polarization is not measured directly as an absolute bulk quantity, but inferred from the charge transferred through the external circuit in response to polarization evolution.

Following this physical picture, consider a parameter-dependent KS Hamiltonian $H(\lambda)=T+V_{\mathrm{KS}}(\lambda)$,
where the KS potential \(V_{\mathrm{KS}}(\lambda)\) depends smoothly on a dimensionless adiabatic parameter \(\lambda\), which represents a slow structural or external perturbation, such as an atomic displacement. As \(\lambda\) evolves continuously from \(\lambda=0\) to \(\lambda=1\), and provided the system remains insulating throughout the process, the derivative of the polarization with respect to \(\lambda\) is well defined in the thermodynamic limit. The total change in polarization can then be expressed as the time-integrated macroscopic current generated during the adiabatic evolution:
\begin{equation}
\Delta \mathbf{P} = \int_{t_i}^{t_j} \mathbf{J}(t) \dd t = \int_{t_i}^{t_j} \frac{\dd \mathbf{P}}{\dd t} \dd t = \int_{t_i}^{t_j} \frac{\partial \mathbf{P}}{\partial \lambda} \frac{\dd \lambda}{\dd t} \dd t = \int_0^1 \frac{\partial \mathbf{P}}{\partial \lambda} \dd\lambda
\end{equation}
Using the conventional definition in Eq.~\ref{eq:convP}, we examine the derivative of the electronic polarization with respect to $\lambda$:
\begin{align}
\frac{\partial \mathbf{P}_\text{elec}}{\partial \lambda} = -\frac{e}{N\Omega} \sum_{\mathbf{k},n \in \text{occ}} \int \mathbf{r} \left( \psi_{\mathbf{k}n}^* \frac{\partial \psi_{\mathbf{k}n}}{\partial \lambda} + \frac{\partial \psi_{\mathbf{k}n}^*}{\partial \lambda} \psi_{\mathbf{k}n} \right) \dd \mathbf{r} \\ \nonumber
= -\frac{e}{N\Omega } \sum_{\mathbf{k},n \in \text{occ}} \left( \mel{\psi_{\mathbf{k}n}}{\mathbf{r}}{\psi'_{\mathbf{k}n}} + \mel{\psi'_{\mathbf{k}n}}{\mathbf{r}}{\psi_{\mathbf{k}n}} \right)
\end{align}
Evaluating this derivative directly is problematic due to the unbounded position operator $\mathbf{r}$. One may use first-order perturbation theory and commutator identities to rewrite the ill-defined position matrix elements in terms of the well-behaved momentum operator. It is also convenient to replace the Bloch wavefunctions $\psi_{\mathbf{k}n}(\mathbf{r})$ with their cell-periodic parts $u_{\mathbf{k}n}(\mathbf{r})$ defined via $\psi_{\mathbf{k}n}(\mathbf{r}) = e^{i\mathbf{k} \cdot \mathbf{r}} u_{\mathbf{k}n}(\mathbf{r})$. After some algebra (see the Supporting Information for details), one obtains
\begin{equation}
\frac{\partial \mathbf{P}_{\text{elec}}}{\partial \lambda} = -\frac{ie}{(2\pi)^3} \sum_{n \in \text{occ}} \int_\text{BZ} \frac{\partial}{\partial \lambda} \mel{u_{\mathbf{k}n}}{\nabla_{\mathbf{k}}}{u_{\mathbf{k}n}} \dd\mathbf{k}
\end{equation}
where \(\int_{\mathrm{BZ}}\) denotes integration over the Brillouin zone (BZ).
Integrating over \(\lambda\) from 0 to 1 gives the total change in electronic polarization:
\begin{equation}
\Delta \mathbf{P}_\text{elec} = -\left. \frac{ie}{(2\pi)^3} \sum_{n \in \text{occ}} \int_\text{BZ} \mel{u_{\mathbf{k}n}}{\nabla_{\mathbf{k}}}{u_{\mathbf{k}n}} \dd\mathbf{k} \right|_{0}^{1} = \mathbf{P}_\text{elec}(\lambda=1) - \mathbf{P}_\text{elec}(\lambda=0)
\label{eq:DeltaPele}
\end{equation}
From this relation, the formal electronic polarization at a given state $\lambda$ is naturally defined as:
\begin{equation}
\mathbf{P}_\text {elec}(\lambda) = -\frac{e}{(2\pi)^3} \sum_{n \in \text{occ}} \int_\text{BZ} i\mel{ u_{\mathbf{k}n}}{\nabla_{\mathbf{k}}}{u_{\mathbf{k}n}} \dd\mathbf{k}
\end{equation}
The integrand $i\mel{ u_{\mathbf{k}n}}{\nabla_{\mathbf{k}}}{u_{\mathbf{k}n}}$ is known as the Berry connection, denoted $\mathcal{A}_n(\mathbf{k})$, associated with the $n$-th occupied Bloch band. This formulation makes explicit that \(\mathbf{P}_{\mathrm{elec}}(\lambda)\) is the Berry phase accumulated by the cell-periodic Bloch states \(u_{\mathbf{k}n}\) over the closed manifold of the Brillouin zone. Therefore, the total formal polarization for an insulating crystal with Hamiltonian $H(\lambda)$ becomes:
\begin{equation}
\mathbf{P}(\lambda) = \mathbf{P}_\text{ion}(\lambda) + \mathbf{P}_\text{elec}(\lambda) = \frac{e}{\Omega} \sum_{s}^{\text{ions}} Z_s \mathbf{r}_s(\lambda) -\frac{e}{(2\pi)^3} \sum_{n \in \text{occ}} \int_\text{BZ} \mathcal{A}_n(\mathbf{k}) \dd\mathbf{k}
\label{eq:totP}
\end{equation}
Because the Berry connection is gauge dependent, a direct integration over the BZ is numerically delicate. In practical implementations, the three-dimensional BZ integral is reformulated as a sum of one-dimensional integrals taken along strings of $\mathbf{k}$-points parallel to the reciprocal lattice vectors~\cite{Jin23p108844}.

\subsection{Adiabatic cyclic loop}
We now show that both the ionic polarization and the electronic polarization defined in Eq.~\eqref{eq:totP} are lattice-valued quantities: they are intrinsically multivalued modulo a polarization quantum.
This fundamental property manifests when analyzing a special class of adiabatic processes in which the Hamiltonian returns to itself: \( H(\lambda = 0) = H(\lambda = 1) \), forming a closed loop in parameter space. Naively, one might expect that a closed adiabatic cycle would result in zero net change in polarization, under the assumption that polarization is a single-valued observable. However, this is generally not the case: both the electronic and ionic contributions to polarization can change by discrete amounts proportional to real-space lattice vectors.

Although the Hamiltonian returns to itself at the end of an adiabatic cycle, the occupied cell-periodic Bloch states (assuming no degeneracy) need not return to itself. They may differ by a band-dependent phase: $\ket{u_{\mathbf{k}n}^{(\lambda=1)}} = e^{i\beta_n(\mathbf{k})} \ket{u_{\mathbf{k}n}^{(\lambda=0)}}$. Because the Bloch states must remain single-valued under translations by a reciprocal lattice vector \(\mathbf{b}_j\), the phase must satisfy $\beta_n(\mathbf{k} + \mathbf{b}_j) - \beta_n(\mathbf{k}) = 2\pi \omega_{n,j}$, with $\omega_{n,j}$ the integer-valued winding number along $\mathbf{b}_j$. Under this gauge transformation, the Berry connection changes as $\mathcal{A}^{(\lambda = 1)}_n(\mathbf{k}) = \mathcal{A}^{(\lambda = 0)}_n(\mathbf{k}) - \nabla_\mathbf{k} \beta_n(\mathbf{k})$. Following Eq.~\ref{eq:DeltaPele}, the change in electronic polarization over the cycle is:
\begin{equation}
\Delta \mathbf{P}_\text{elec} = -\frac{e}{(2\pi)^3} \sum_{n \in \text{occ}} \int_\text{BZ} \nabla_{\mathbf{k}} \beta_n(\mathbf{k}) \dd\mathbf{k} = -\frac{e}{(2\pi)^3} \sum_{n \in \text{occ}} \oint_{\partial\text{BZ}} \beta_n(\mathbf{k}) \dd\mathbf{S} = -\frac{e}{\Omega} \sum_{n \in \text{occ}} \mathbf{R}_n,
\end{equation}
where \(\mathbf{R}_n = \sum_{j=1}^3 \omega_{n,j}\mathbf{a}_j\) is a lattice vector, with \(\mathbf{a}_j\) (\(j=1,2,3\)) denoting the primitive lattice vectors. The divergence theorem is used to convert the BZ integral as surface integral over BZ boundary (see detailed derivation in Supporting Information).
Thus, the change in electronic polarization over an adiabatic cycle is rigorously quantized:
\begin{equation}
\Delta \mathbf{P}_\text{elec} = -\frac{e}{\Omega}\mathbf{R}, \qquad \mathbf{R} \in \text{Bravais lattice}.
\end{equation}
Because \(\Delta \mathbf{P}_\text{elec}\) is multivalued, it follows immediately that \(\mathbf{P}_\text{elec}\) itself is also multivalued, defined modulo \(e\mathbf{R}/\Omega\).

For the ionic contribution, the ions follow continuous trajectories \(\mathbf{r}_s(\lambda)\) during the same adiabatic cycle. Although the crystal structure at \(\lambda=1\) is identical to that at \(\lambda=0\), individual ions need not return to their original positions; they may instead end at symmetry-equivalent sites displaced by lattice vectors \(\mathbf{R}_s\). The resulting change in ionic polarization is therefore
\begin{equation}
\Delta \mathbf{P}_\text{ion} = \int_0^1 \frac{\dd \mathbf{P}_\text{ion}}{\dd \lambda} \dd\lambda = \frac{e}{\Omega} \sum_s^{\text{ions}} Z_s [\mathbf{r}_s(1) - \mathbf{r}_s(0)] = \frac{e}{\Omega} \sum_s^{\text{ions}} Z_s \mathbf{R}_s
\end{equation}
Since each \(Z_s\) is an integer and each \(\mathbf{R}_s\) is a Bravais lattice vector, $\sum_s^{\text{ions}} Z_s \mathbf{R}_s$ is also a Bravais lattice vector, denoted \(\mathbf{R}'\). Hence,
\begin{equation}
\Delta \mathbf{P}_\text{ion} = \frac{e}{\Omega} \mathbf{R}', \qquad  \mathbf{R}' \in \text{Bravais lattice}.
\end{equation}

Therefore, a central result of the MTP is that the formal total polarization $\mathbf{P}$ of a crystal is multivalued  modulo $\mathbf{Q}$:
\begin{equation}
\mathbf{P} \in \left\{ \mathbf{P}_0 + \sum_{i=1}^3 n_i \mathbf{Q}_i \;\middle|\; \mathbf{Q}_i = \frac{e}{\Omega} \mathbf{a}_i, \; n_i \in \mathbb{Z} \right\}.
\label{eq:key1}
\end{equation}
Here, $\mathbf{P}_0$ is the principal value that depends on the specific atomic positions and species in the unit cell, and $\mathbf{Q}_i$ represent the smallest possible change in polarization (the polarization quantum) along each lattice direction, corresponding to the transport of a single elementary charge by one primitive lattice vector $\mathbf{a}_i$. 

\subsection{Classical point charge model}

The multivalued nature of polarization is not exclusively a quantum phenomenon. In fact, this ambiguity can also be understood within a purely classical framework using a model of point charges. Consider a two-dimensional (2D) crystal composed of classical point particles with integer charges $Z_i e$. For simplicity, we analyze a square lattice containing alternating cations ($Z_i=+1$) and anions ($Z_i=-1$) with lattice vectors $\mathbf{a}_1$ and $\mathbf{a}_2$. 
The formal polarization $\mathbf{P}$ in 2D is defined as the dipole moment per unit cell area $S$, $\mathbf{P} = \frac{e}{S} \sum_{i} Z_i \mathbf{r}_i$, where the sum runs over all point charges within a chosen primitive cell, and \(\mathbf{r}_i\) denotes the position of the \(i\)-th charge measured relative to the chosen unit-cell origin.
This expression is structurally identical to the ionic polarization defined earlier; the only difference is that the charges \(Z_i\) may now be either positive or negative.  

To evaluate $\mathbf{P}$, one must define a specific unit cell (a basis). However, in a periodic system, this choice is not unique. 
As illustrated in Figure \ref{fig:classical_ambiguity}, we can construct equally valid unit cells that describe the same 2D crystal but yield different polarization values. In one choice of unit cell (Choice A), the anion at the origin is paired with an cation at relative position \(\mathbf{r}\), giving a dipole moment \(\mathbf{d}_A = e\mathbf{r}\). In an alternative choice (Choice B), the unit-cell boundary is shifted so that the anion is paired with a cation at position \(\mathbf{r}-\mathbf{R}\). The dipole moment then becomes \(\mathbf{d}_B = -e(\mathbf{r} - \mathbf{R})\). The resulting difference in polarization is $\mathbf{P}_B - \mathbf{P}_A = -{e \mathbf{R}}/{S}$.
Since $\mathbf{R}=\mathbf{a}_1+\mathbf{a}_2$ is a lattice vector, the difference is exactly the polarization quantum in 2D. This example shows that even in a classical point-charge model, the formal polarization is multivalued modulo \(\mathbf{Q}\).

\begin{figure}[h]
    \centering
    \includegraphics[width=\textwidth]{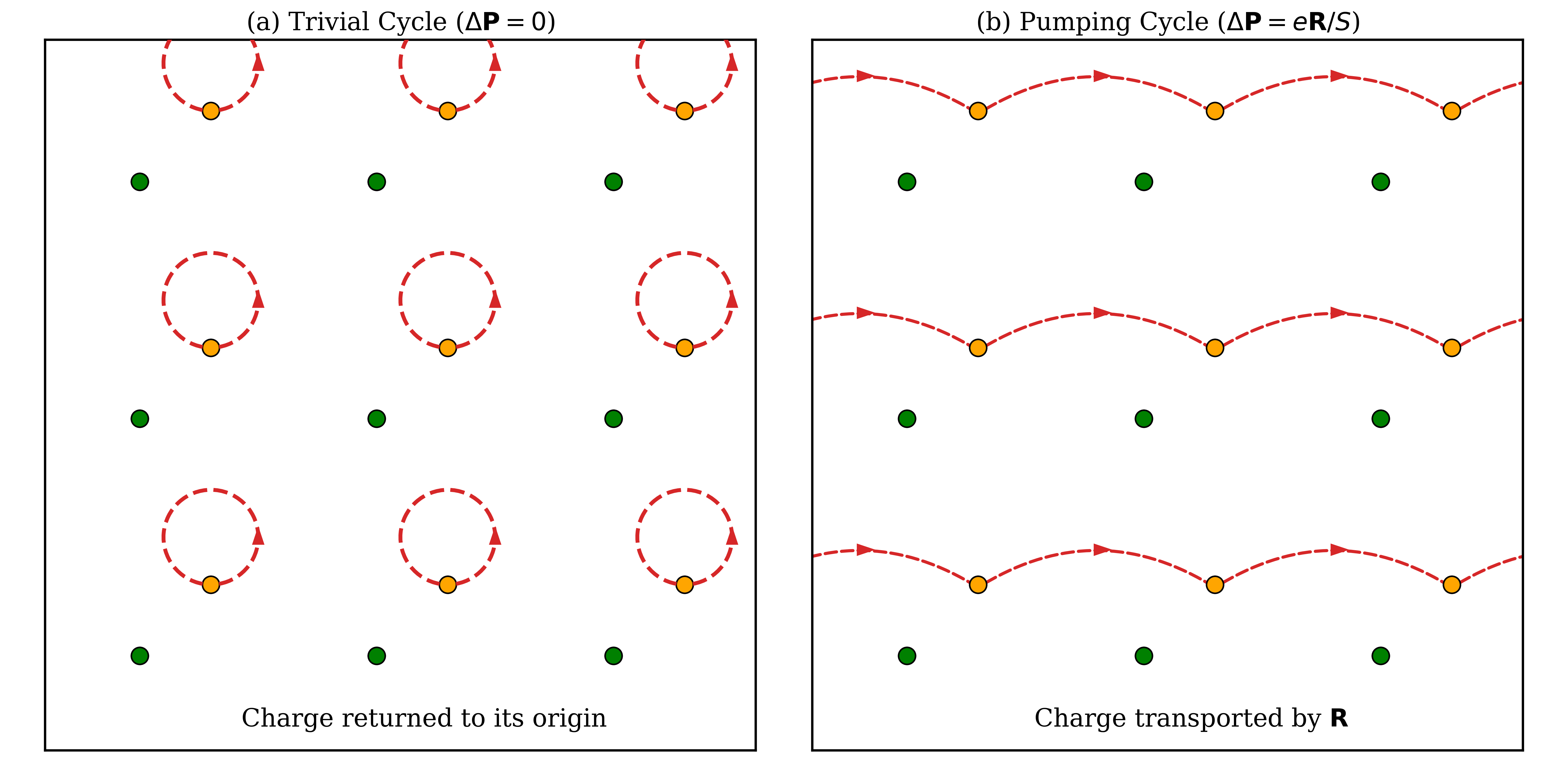}
\caption{Two distinct cyclic evolutions in a two-dimensional crystal of classical point charges. (a) Trivial cycle: the anions perform  closed loops around their equilibrium positions, yielding zero net current. (b) Pumping cycle: the anions migrate into neighboring unit cells. Although the bulk crystal appears identical at the beginning and end of the cycle, a net charge is transported by one lattice vector \(\mathbf{R}\), resulting in a polarization change $\Delta \mathbf{P}$ of \(e\mathbf{R}/S\).}
    \label{fig:cyclic_loops}
\end{figure}

We can also understand the cyclic adiabatic process using this classical model. Two types of cycles can be distinguished based on whether there is net charge transport.
In a \textit{trivial} cycle [Fig.~\ref{fig:cyclic_loops}(a)], the cations move along a closed loop in real space and return to their exact original positions. As the net displacement of charge is zero, the trivial cycle has $\Delta \mathbf{P} = 0$. In a \textit{nontrivial} cycle [Fig.~\ref{fig:cyclic_loops}(b)], the cations are translated by one lattice vector \(\mathbf{a}_1\) (over a time \(t_0\)). Although the crystal structure is restored at the end of the cycle, a net charge has been transported across the system. The corresponding 2D current density, defined as the current per unit transverse length, is $\mathbf{j} = \frac{e}{t_0|\mathbf{a}_2|}\,\hat{\mathbf{a}}_1
= \frac{e\mathbf{a}_1}{S\,t_0}$.  Integrating over time yields $\Delta \mathbf{P} = {e\mathbf{a}_1}/{S} =\mathbf{Q}_1$. 
This already shows that, if such a nontrivial cycle can be realized, it can produce a finite polarization change between two otherwise identical crystal structures. This insight has important implications for understanding FQFEs.

\subsection{Periodic boundary conditions}
At first sight, cyclic adiabatic loops in which ions are displaced by lattice vectors may seem artificial. After all, in a finite sample an ion shifted by a lattice vector will leave the crystal rather than return to an equivalent position as assuming periodic boundary conditions (PBCs). This raises a natural question: why does the MTP formulate polarization using PBCs?

Prior to the MTP, defining polarization via the dipole moment per unit volume led to ambiguities in periodic systems as explained above, casting doubts about whether it was a well-defined \textit{bulk} observable. 
By a bulk property, we mean a quantity defined in the thermodynamic limit and determined entirely by the translationally invariant interior of the crystal, independent of the microscopic details of its boundaries or surface termination. In this sense, the formal polarization in the MTP is not defined by examining a particular finite sample with edges, but by characterizing the infinite periodic solid described by Bloch states. PBCs are therefore not merely a mathematical convenience; they isolate the intrinsic bulk response of the crystal from extrinsic surface contributions.

When PBCs are lifted, as in any finite system, the total polarization recovers the single-valued definition of $\mathbf{P} = \mathbf{d}/{V} = {V}^{-1} \int \mathbf{r} \rho(\mathbf{r}) \dd\mathbf{r}$, which includes both bulk and surface contributions because $\mathbf{r}$ is bounded~\cite{Huang26p020701}. We emphasize that the multivaluedness of the formal polarization remains physically relevant even without PBCs, as expressed by the surface-charge theorem~\cite{Vanderbilt93p4442}. This discussion assumes a gapped (insulating) surface termination, for which the surface charge density $\sigma_\text{surf}$ is well defined as the projection of the bulk polarization $\mathbf{P}$ onto the surface normal unit vector $\hat{\mathbf{n}}$, such that $\sigma_\text{surf} = \mathbf{P}\cdot \hat{\mathbf{n}}$. The projection of a polarization quantum $\mathbf{Q}$ (where we omit the direction index $i$ for simplicity) satisfies $\frac{e}{\Omega}\mathbf{a}\cdot\hat{\mathbf{n}} = \frac{e}{S}$, where $S$ is the surface unit-cell area. Without specifying the surface termination, the surface charge density  is restricted to a discrete set fixed by the bulk polarization:
\begin{equation}
    \sigma_\text{surf} \in \left\{ \mathbf{P}_0\cdot\hat{\mathbf{n}} + m\frac{e}{S} \;\middle|\; m \in \mathbb{Z} \right\}.
\end{equation}
Once a particular termination $i$ is specified, the physical surface charge density becomes single-valued:
\begin{equation}
    \sigma_\text{surf}^{(i)}= \mathbf{P}_0\cdot  \hat{\vb{n}} + m_i\frac{e}{S},
\end{equation}
where $m_i$ depends on the atomic and electronic structure of that surface. The surface charge theorem therefore shows that boundary observables are constrained by the bulk polarization topology of the insulating crystal.

For a concrete illustration of the bulk--boundary correspondence, see Kudin, Car, and Resta~\cite{Kudin05p13}. Studying quasi-one-dimensional push--pull polymers, such as functionalized trans-polyacetylene chains, they linked the bulk formal polarization from Berry-phase theory to the dipole moment of finite oligomers with broken translational symmetry at the ends. Their key result is that the bulk fixes a quantized set of allowed surface charges, while the actual surface charge, and thus the dipole moment, depends on the surface termination. This provides a clear first-principles demonstration of the surface charge theorem and the bulk--boundary correspondence.

\subsection{Symmetry of formal polarization}

Since formal polarization is a multivalued quantity, the symmetry constraints that apply to it differ fundamentally from those applied to conventional vector quantities. For standard vectors, symmetry arguments dictate that in a centrosymmetric crystal, the vector must vanish.
However, formal polarization is not a single vector but rather a lattice of values, defined modulo the polarization quantum. In this context, the requirement of inversion symmetry is satisfied not when the polarization itself vanishes, but when the entire lattice of allowed polarization values maps onto itself under the symmetry operation. 
In a centrosymmetric crystal, there are two distinct ways this condition can be satisfied. In the first case, the polarization lattice is centered at the origin, such that inversion symmetry maps each value to its negative, and the lattice remains invariant. For example, in a simple 1D system where the polarization quantum is 1, the allowed values could be $\{..., -2, -1, 0, +1, +2,...\}$, and inversion symmetry maps each point to another within the same set. In the second case, the lattice straddles the origin, for instance, $\{..., -\frac{3}{2}, -\frac{1}{2}, +\frac{1}{2}, +\frac{3}{2},...\}$. Here, inversion symmetry still maps the set onto itself, even though none of the individual values are zero. Figure 2 illustrates these two scenarios in a 2D setting.
A real material example of the second case is centrosymmetric KNbO$_3$~\cite{Resta93p1010}, for which first-principles calculations using the Berry-phase approach yield a formal polarization of 51 $\mu$C/cm$^2$, exactly half of the polarization quantum of $102$ $\mu$C/cm$^2$. 
This non-zero value does not violate Neumann’s Principle, as it only corresponds to one element of a polarization lattice.

This concept extends beyond inversion to arbitrary symmetry operations. As formalized by the generalized Neumann's principle~\cite{Pang25p116402}, a symmetry operation does not need to keep a given value of the formal polarization exactly the same; it only needs to change it by a polarization quantum, so that the new value remains on the same polarization lattice. Specifically, under a crystallographic symmetry operation $\mathcal{R}$, a branch of the polarization $\mathbf{p}$ must satisfy
\begin{equation}
   \mathcal{R}\mathbf{p} -  \mathbf{p} = (\mathcal{R}-\mathcal{I})\mathbf{p} = \mathbf{Q},
    \label{eq:generalized_neumann}
\end{equation}
where $\mathcal{I}$ is the identity operation and $\mathbf{Q}$ is a polarization quantum. Equivalently, the polarization lattice as a whole is invariant under the symmetry.

\subsection{Effective polarization}
Resta and Vanderbilt introduced the \textit{effective polarization} $\mathbf{P}_{\mathrm{eff}}$, which is consistent with experimental $P$--$E$ hysteresis measurements~\cite{Resta94p899,KingSmith93p1651}. It is defined as the change in polarization along an adiabatic path from a reference structure $(\lambda=0)$ to the ferroelectric state $(\lambda=1)$:
\begin{equation}
    \mathbf{P}_{\mathrm{eff}}=\Delta \mathbf{P}=\mathbf{P}(\lambda=1)-\mathbf{P}(\lambda=0).
\end{equation}
The reference structure is usually the high-symmetry centrosymmetric phase. It is assumed to be adiabatically connected to the ferroelectric state by a collective ionic displacement, often associated with a soft mode, while the system remains insulating throughout the path. In practice, the polarization is usually computed for a sequence of intermediate structures, often within density functional theory, so that its evolution can be tracked continuously on the same branch. The resulting $\mathbf{P}_{\mathrm{eff}}$ can therefore be compared directly with the measured polarization.

The theoretical determination of \(\mathbf{P}_{\mathrm{eff}}\) involves subtleties beyond simple symmetry arguments, especially in the choice of the high-symmetry reference phase. In practice, this reference is often selected based on physical intuition, typically as the nearest high-symmetry structure related to the ferroelectric ground state through a group--subgroup relation. This choice works well for many conventional ferroelectrics such as PbTiO\(_3\) and BaTiO\(_3\). Yet this criterion is not sufficient in general. For meaningful comparison with polarization values measured from \(P\)--\(E\) hysteresis loops, the reference structure must not only be symmetry allowed, but also correspond to the actual intermediate configuration accessed during field-driven switching.

A notable example is provided by the wurtzite III-nitrides (GaN, AlN, and InN; Fig.~\ref{fig:AlN}(a)), for which the zinc-blende (ZB) structure (Fig.~\ref{fig:AlN}(c)) has often been used as a high-symmetry reference because of its structural similarity~\cite{Bernardini97pR10024,Bernardini01p193201,Bechstedt00p8003}. This choice, however, introduces an important error. Although the ZB structure belongs to the nonpolar space group \(F\bar{4}3m\), it carries a nonzero formal polarization along \([111]\), as allowed by the multivalued nature of polarization. As shown by Dreyer \textit{et al.}, using the ZB structure as the reference therefore introduces a material-dependent correction to \(\mathbf{P}_{\mathrm{eff}}\) in wurtzite systems~\cite{Dreyer16p21038}. A more appropriate reference is the layered hexagonal structure (space group \(P6_3/mmc\); Fig.~\ref{fig:AlN}(b)), which is centrosymmetric and has vanishing formal polarization by construction.

Moreover, the hexagonal phase can be connected to wurtzite through a path that is plausibly driven by an electric field during switching. As shown in Figs.~\ref{fig:AlN}(a) and \ref{fig:AlN}(b), a downward electric field can drive the positively charged Al cations toward the N planes, thereby reversing the polarization. By contrast, the same field is unlikely to drive the wurtzite phase toward ZB phase [Fig.~\ref{fig:AlN}(c)], since that transformation would require the Al cations to move against the field direction. Using the hexagonal phase as the reference gives $\mathbf{P}_{\mathrm{eff}} = 1.35$~C/m$^2$ for wurtzite AlN, consistent with measurements in Sc-doped AlN~\cite{Fichtner19p114103}. Using zinc blende instead yields a negligible value of $-0.090$~C/m$^2$.

Another important requirement for defining $\mathbf{P}_{\mathrm{eff}}$~\cite{Hohenberg64pB864, Kohn65pA1133} is that the switching path itself must be physically accessible under an electric field. In practice, switching barriers are often estimated using DFT-based nudged elastic band (NEB) calculations~\cite{Henkelman00p9901, Henkelman00p9978} to identify a minimum-energy path. However, such paths are usually constructed \textit{ad hoc} and need not correspond to field-activated switching. The existence of a continuous geometric path between two polar states does not guarantee that the path can be driven by an electric field.
AB-stacked bilayer $h$-BN illustrates this point clearly~\cite{Yasuda21p1458, ViznerStern21p1462, woods21p347}. The AB and BA configurations have opposite out-of-plane polarizations, yet both belong to the same polar space group $P3m1$ and preserve out-of-plane threefold rotational ($C_3$) symmetry. Several DFT-based NEB studies have identified an in-plane sliding path between them with a relatively low energy barrier~\cite{Li17p6382}. However, finite-field DFT calculations by Ke \textit{et al.} showed that the in-plane forces induced by an out-of-plane electric field vanish exactly in both AB and BA single-domain configurations~\cite{ke25p046201}. This follows from the exact cancellation of the off-diagonal components of the Born effective charge tensor required by $C_3$ symmetry. Consequently, the interlayer sliding motion needed to reverse the polarization is symmetry-forbidden in a single-domain state and cannot be activated by an electric field. This example shows that defining $\mathbf{P}_{\mathrm{eff}}$ requires not only a continuous path between two polar states, but also one that is physically accessible under an applied electric field.

\section{Fractional Quantum Ferroelectricity}

\subsection{Fractionally Quantized Polarization and Oxidation States }
The concept of FQFEs was recently introduced to describe materials in which polarization changes arise from large atomic displacements comparable to a lattice vector, in contrast to the small displacements typical of conventional ferroelectrics such as PbTiO$_3$~\cite{Lente01p5093, Ghosez00p2767}. High-throughput DFT studies have identified more than 200 candidate materials, including bulk AlAgS$_2$~\cite{Range76p311} and monolayer HgI$_2$~\cite{Jeffrey67p396, Schwarzenbach07p828}. A defining, and initially counterintuitive, feature of FQFEs is that materials in nonpolar space groups can still exhibit polarization changes along an adiabatic path connecting symmetry-equivalent structures.

In early reports, the appearance of polarization in nonpolar point groups, such as the $T_d$ symmetry of zinc-blende AgBr~\cite{Wu18pe1365, Gao19p1106}, was interpreted as a possible violation of Neumann's principle. As discussed in Sec.~II, the MTP shows that polarization is a lattice-valued quantity. For a crystal with a nonpolar point group, all branches of the polarization lattice may still be nonzero (as in KNbO$_3$). Consequently, an adiabatic path connecting symmetry-equivalent structures, much like a cyclic adiabatic loop, can produce a nonzero polarization change.

The large polarization changes in FQFEs can be understood through the topological definition of oxidation state within the MTP. Jiang \textit{et al.} showed that the oxidation state of an ion in an insulator can be defined as a topological invariant~\cite{Jiang12p166403}. As discussed in Sec.~II.B, translating a sublattice by a full primitive lattice vector $\mathbf{a}$ returns the Hamiltonian to itself [Fig.~\ref{fig:classical_ambiguity}(b)]. The associated change in polarization is therefore strictly quantized:
\begin{equation}
\Delta \mathbf{P} = N\frac{e\mathbf{a}}{\Omega} = N\mathbf{Q},
\end{equation}
where $N$ is the oxidation state, or effective charge, of the ion. It is given by the integer difference between the nuclear charge and the number of Wannier centers transported with the nucleus during the lattice translation.

FQFEs provide a direct physical realization of this topology. In these systems, switching involves displacing a sublattice by only a fraction of a lattice vector, $\mathbf{u}=\mathbf{a}/m$, where $m>1$ is an integer. After $m$ such fractional displacements, one reconstructs the usual closed path corresponding to a full lattice translation. The polarization change for a single FQFE switching event is then
\begin{equation}
\Delta \mathbf{P}_{\rm FQFE}=\frac{\Delta \mathbf{P}}{m}=\frac{N}{m}\mathbf{Q}.
\end{equation}
Following the classification of Yu \textit{et al.}~\cite{Yu25p016801}, systems with fractional $N/m$ are termed Type-I FQFEs. For example, displacing an ion with oxidation state $+1$ by half of a primitive lattice vector $(m=2)$ gives $\Delta \mathbf{P}_{\rm FQFE}=\mathbf{Q}/2$ [Fig.~\ref{fig:FQFE_Type}(a)]. By contrast, if $N/m$ is an integer, such as $N=+2$ and $m=2$, the system is classified as a Type-II FQFE [Fig.~\ref{fig:FQFE_Type}(b)]. This relation also provides a direct way to extract the oxidation state $N$ from fractional switching data when $\Delta \mathbf{P}_{\rm FQFE}$, $m$, and $\mathbf{Q}$ are known.

Monolayer $\alpha$-In$_2$Se$_3$ is often cited as a prototypical system for exploring the possible emergence of FQFE. Despite its $C_{3v}$ point group symmetry, which under conventional symmetry analysis strictly forbids the existence of an intrinsic in-plane polarization vector, several experiments have reported evidence of in-plane switchable polarization~\cite{Ding17p14956, Soleimani20p22688, Zhou17p5508, Cui18p1253, Xue18p1803738, Xue18p4976, Xiao18p227601}. 
Figure~\ref{fig:FQFE_Type}(c) shows the evolution of the in-plane polarization along a continuous path connecting two symmetry-equivalent $\alpha$-In$_2$Se$_3$ structures ($L_1$ and $L_2$), as calculated using DFT-based Berry phase approach. The formal polarization exhibits a change  from $\mathbf{Q}/3$ to $-\mathbf{Q}/3$, which results in a net polarization change of $\Delta \mathbf{P}_{\rm FQFE} = -2\mathbf{Q}/3$. Since the switching event corresponds to a fractional displacement of $\mathbf{u}=\mathbf{R}/{3}$ (i.e., $m=3$), this implies an oxidation state of $N=-2$ for the Se ion, consistent with its nominal charge.

\subsection{Is $\alpha$-In$_2$Se$_3$ really a FQFE?}
From the analysis above, it becomes clear that a change in polarization within a crystal belonging to a nonpolar space group does not violate any fundamental symmetry constraints when viewed through the lens of the MTP. In this context, experimental reports of in-plane polarization in monolayer $\alpha$-In$_2$Se$_3$ have been interpreted as possible evidence for FQFE, because its out-of-plane threefold rotational symmetry should forbid any nonzero in-plane vector property. However, we urge caution in definitively classifying $\alpha$-In$_2$Se$_3$ as an FQFE candidate, for the following reasons.

First, nearly all claims of in-plane polarization in $\alpha$-In$_2$Se$_3$ are based on lateral piezoresponse force microscopy (PFM) measurements performed in Dual AC Resonance Tracking (DART) mode~\cite{Cui18p1253, Xue19p1901300, Xue18p4976, Li20p2000061, Wang20p2004609}. However, it is well known that out-of-plane and in-plane PFM signals can be strongly mixed, which can lead to misinterpretation. In a recent study, two probes with different force constants were compared: a stiff probe [$k \approx 2.8$ N/m, Fig.~\ref{fig:InSe_FQFE}(a)] and a soft probe [$k \approx 0.2$ N/m, Fig.~\ref{fig:InSe_FQFE}(b)]. Both probes showed strong out-of-plane piezoresponse and a clear $180^\circ$ phase contrast between upward and downward domains. In contrast, the in-plane phase signal measured with the soft probe was only marginal, suggesting that the apparent in-plane contrast observed with the stiff probe arose from crosstalk from the out-of-plane response.

Second, angle-resolved lateral PFM, in which the sample is rotated with respect to the cantilever, also argues against intrinsic in-plane piezoresponse. In particular, $\alpha$-In$_2$Se$_3$ samples with prewritten box-in-box domain patterns show no azimuthal dependence in either the lateral PFM amplitude or phase [Fig.~\ref{fig:InSe_FQFE}(c)], which is in stark contrast to the expected periodic modulation observed in reference materials like NbOI$_2$ [Fig.~\ref{fig:InSe_FQFE}(d)]~\cite{Bai24p26103}.

Third, $\alpha$-In$_2$Se$_3$ is known to be leaky and can host a competing ferroelectric phase with in-plane polarization. As a result, even measured polarization--electric field hysteresis loops may be affected by ionic conduction, defect migration, or field-induced phase transitions. These effects complicate any direct identification of intrinsic bulk ferroelectricity.

Further insight was provided by Bai \textit{et al.}, who carried out large-scale molecular dynamics simulations using deep-learning-based interatomic potentials~\cite{Bai24p26103}. They found that single-domain monolayer $\alpha$-In$_2$Se$_3$ is difficult to switch, even under very strong electric fields, because the nucleation barrier is prohibitively high. Although NEB calculations suggest a low minimum-energy switching path, the dominant energy cost comes from the interfacial energy of the nucleated domain, not from the local reversal itself.

To explain the apparent switching reported experimentally, Bai \textit{et al.} proposed a domain-wall-mediated mechanism~\cite{Bai24p26103}. Although the single-domain bulk preserves $C_{3v}$ symmetry and therefore lacks in-plane ferroelectricity, one-dimensional domain walls can locally break this symmetry [Fig.~\ref{fig:InSe_FQFE}(e)]. These walls carry localized polarization charge due to formal polarization discontinuities, allowing in-plane electric fields to move them through creep and avalanche dynamics [Fig.~\ref{fig:InSe_FQFE}(f)]. At the atomic scale, this motion is driven by a unique "stone-skipping-like" mechanism rather than the conventional line-by-line progression. The observed switching behavior may therefore originate from domain-wall motion rather than conventional bulk ferroelectric switching.
These results show that classifying $\alpha$-In$_2$Se$_3$ as an FQFE remains challenging. A reliable conclusion requires careful treatment of experimental artifacts and realistic modeling of domain-wall dynamics, as well as possible electric-field-driven ion transport, which is central to FQFE behavior.

\section{Ionic Conductor Ferroelectricity}

\subsection{Giant Quantized Polarization Change}
Moving beyond FQFEs, which involves ion displacements that are fractional multiples of the primitive lattice vectors within a unit cell, the concept of ICFEs further expands the boundaries of conventional ferroelectricity~\cite{Resta92p51, Resta94p899, Wang24p3885, Lu23p281, Zhong19p373, Yanagisawa24p1476, wang22p9552, Ren18p35361}. 
First proposed through DFT studies, ICFEs exhibit ion displacements that often exceed typical covalent bond lengths, frequently involving cross-unit-cell migrations.
Similar to FQFEs, ICFEs also allow the emergence of polarization in materials with nonpolar point groups (e.g., $C_{2h}$ in Mg$_{0.5}$CrS$_2$). A defining characteristic of an ICFE is the ultra-long ionic displacement, which can extend over multiple lattice constants, especially in the presence of vacancies, leading to quantized, polarization change.
Representative examples include 2D layered materials such as CuCrS$_2$~\cite{Zhong19p373, wang22p9552} and InSe~\cite{Bandurin16p223, Ren18p35361}, as well as superionic conductors like Na$_4$SnS$_4$~\cite{Gao21p12362, wang22p9552} and KSnS$_4$~\cite{Xiong20p100281, wang22p9552}. 
Experimental studies have reported phenomena consistent with theoretical predictions. 
For instance, in CuCrS$_2$, the migration of intercalated Cu ions across different symmetry-equivalent sites (States I, II, and III) results in giant, stepwise changes in in-plane polarization [Fig.~\ref{fig:ICFE}(a)], showing room-temperature ferroelectric-like switching coexisting with ionic conductivity. Similarly, the organic-inorganic hybrid (CETM)$_2$InCl$_5 \cdot$H$_2$O exhibits an enormous polarization of up to 31,654\,$\mu$C/cm$^2$~\cite{Henkelman00p9978}, attributed to field-driven proton migration along specific pathways [Fig.~\ref{fig:ICFE}(b)].

Following the discussion of the MTP in Sec.~II and the topological definition of oxidation states in Sec.~III.A, the large polarization changes in ICFE materials can be understood as arising from adiabatic charge pumping. When a mobile ion migrates to a symmetry-equivalent site in a neighboring unit cell, the system undergoes an adiabatic cycle and the polarization changes by a quantized amount. Specifically,
$\Delta \mathbf{P} = eN\mathbf{D}/{\Omega}$, 
where $N$ is the oxidation state of the migrating ion and $\mathbf{D}$ is its displacement vector. In DFT calculations, $\mathbf{D}$ is often taken to be an integer multiple of a primitive lattice vector, $\mathbf{D}=m\mathbf{a}$ with $m \in \mathbb{Z}$. The apparently ``giant'' polarization values reported for ICFEs therefore reflect the geometric and topological consequence of transporting quantized charge over discrete lattice distances.

\subsection{Challenges in Practical Implementation}

While the theoretical basis of ICFEs is well justified, its practical use in devices raises important challenges, especially regarding switching reversibility and interface stability. In idealized DFT-based NEB calculations, ion migration is treated as a coherent collective process in which mobile ions move to symmetry-equivalent sites in neighboring unit cells. Under PBCs, this motion occurs without net ion accumulation or depletion.

The realization of a stable macroscopic polarization in these layered ionic conductors, such as $\text{KSnS}_4$ [Fig.~\ref{fig:ICFE}(c)], depends heavily on the thermodynamic behavior of intrinsic point defects. Specifically, mobile ion vacancies within the conduction channel can either distribute evenly due to Coulomb repulsion (the nonpolar ``D state") or aggregate at one side (the polar ``A state") [Fig.~\ref{fig:ICFE}(d)]. For an ionic conductor to exhibit practical ferroelectric-like properties, the polar A state must be the energetically favorable ground state; otherwise, the induced polarization will not be stably retained.

In real devices, however, the finite size of the material changes the situation. Once an electric field is strong enough to drive ions across one unit cell, there is in principle no intrinsic limit to how far they can continue to migrate. If ions are blocked at an interface, they can accumulate near one electrode and be depleted near the other. This ionic redistribution alters the local electrostatics, including the screening charge at the electrodes, and may generate an external current that resembles ferroelectric switching~\cite{Huang24p6683}. At the same time, ion accumulation or vacancy buildup can destabilize the interface and degrade the device. It is also unclear whether ions trapped in a likely disordered interfacial region can be driven back reversibly by reversing the field.

If ions are not blocked and can pass through the interface, the active layer continuously loses ionic species. Such irreversible migration would rapidly degrade the material and eventually cause device failure. Thus, although ICFE materials can produce large polarization currents, using them in conventional structures such as metal--ferroelectric--metal capacitors~\cite{Wan19p1808606, Garcia14p4289, Junquera03p506, Xue21p7291, Gerra06p107603, Yang22p1422} requires careful control of ion reversibility and robust interface engineering to prevent degradation and enable repeatable operation~\cite{Wang12p247601, Zhang17p1703543, Lin21p7}.

\section{Make usage of formal polarization}
We emphasize that the emerging concepts of FQFEs and ICFEs are fully consistent with the MTP. Both arise naturally from the Berry-phase theory of formal polarization and represent unconventional routes to obtain effective polarization through long-range ionic motion.
Returning to the original spirit of Valasek's definition, a defining feature of ferroelectricity is the reproducible measurement of a switching current under an applied electric field. For both FQFE and ICFE systems, the key open question is therefore whether such field-driven switching is truly reversible under experimentally relevant conditions.

One promising application of the giant polarization in FQFEs lies in interface physics. As shown by Stengel and Vanderbilt for oxide heterostructures, the macroscopic bound charge density at a coherent interface between two insulators is determined by the discontinuity in their bulk formal polarizations~\cite{Stengel09p241103}:
\begin{equation}
    \sigma_{\text{bound}} = (\mathbf{P}_1 - \mathbf{P}_2) \cdot \hat{n},
\end{equation}
where $\hat{n}$ is the interface normal. 
As a result, interfaces between materials with mismatched formal polarizations can naturally host emergent phenomena, such as two-dimensional electron gases or pronounced lattice distortions, which act to compensate the adiabatic discontinuity~\cite{Pang25p18697}. Moreover, such discontinuities in formal polarization may also be generated dynamically through interlayer sliding in two-dimensional materials~\cite{Lu25p21559}. From this perspective, the most promising practical route toward exploiting FQFEs may be the intentional engineering of interfaces and domain walls that host formal-polarization discontinuities. These discontinuities generate bound charge and, if directly coupled to external electric fields, may be manipulated in a controlled fashion.

A particularly promising application of ICFE materials is neuromorphic computing, especially in electrochemical ionic synapses (EIS)~\cite{Huang23p2205169, Onen22p539}. Recent studies of ferroelectric HfO$_2$ have shown that a unidirectional electric bias can drive long-range oxygen migration with high mobility [Fig.~\ref{fig:ICFE}(e)]. This enables HfO$_2$ to function as the electrolyte layer in EIS devices, which operate analogously to nanoscale batteries for in-memory computing. In a typical EIS structure, a ferroelectric $Pca2_1$ HfO$_2$ layer connects an oxygen reservoir to a channel layer whose conductivity depends on the local oxygen-vacancy concentration. Under an applied bias, oxygen ions migrate through the ferroelectric electrolyte and modulate the vacancy concentration in the channel [Fig.~\ref{fig:ICFE}(f)], thereby tuning its conductance in a stepwise fashion and enabling analog synaptic weights. In this sense, ferroelectric HfO$_2$ can be viewed as an ICFE. More broadly, the high field-driven ionic mobility of ICFE materials makes them attractive for scalable, high-speed EIS devices.

\section{Conclusion}
Since the modern theory of polarization was established in the 1990s, it has been remarkable to see how the concept of formal polarization continues to guide the discovery of new material classes. Although ferroelectricity has been studied for more than a century, the recent proposal of FQFEs and ICFEs shows that formal polarization, once regarded as abstract, or even ``formally useless'', may in fact point toward new and tangible functionalities. Future progress will require a deeper understanding of the dynamics of these systems, especially the role of interfaces and domain walls. A major open question concerns their functional response: unlike conventional ferroelectrics in polar space groups, which exhibit intrinsic piezoelectric and pyroelectric effects tied to soft-mode distortions, it remains unclear whether FQFE and ICFE materials display the same canonical responses or instead operate through fundamentally different electromechanical mechanisms, possibly accessible only through interface engineering.

To better distinguish these mechanisms, we also suggest the term \textit{topological ionics}. This term refers to materials in which polarization changes arise from ionic motion between topologically distinct branches of the formal polarization lattice, in accordance with the generalized Neumann's principle. If such materials can be controlled reliably, they may enable a new class of nonvolatile memories that combine the robustness of ionic conductors with the switchability of dipolar systems.

\begin{acknowledgments}
We acknowledge the supports from Zhejiang Provincial Natural Science Foundation of China (LR25A040004). The computational resource is provided by Westlake HPC Center. We gratefully acknowledge Zhuang Qian from Westlake University for helpful conversations. 
\end{acknowledgments}

\bibliography{SL.bib}

\clearpage
\newpage

\clearpage
\newpage
\clearpage
\newpage
\begin{figure}[htbp]
    \centering
    \includegraphics[width=0.8\textwidth]{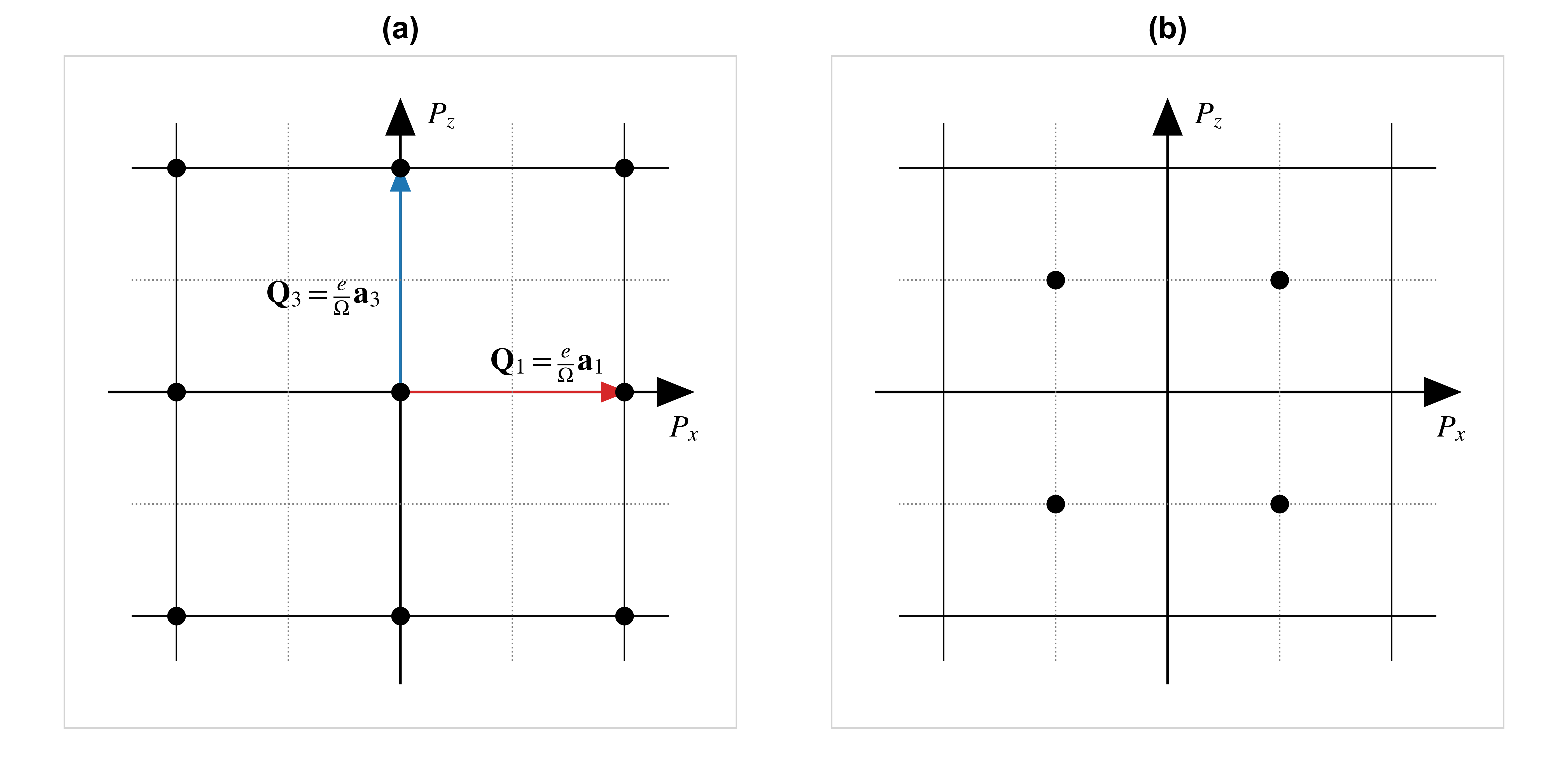} 
\caption{Schematic illustration of the two polarization lattices compatible with inversion symmetry in a centrosymmetric crystal. (a) Case A: the polarization lattice contains the origin (\(\mathbf{P}=0\)), which is invariant under inversion. (b) Case B: the polarization lattice is shifted relative to the origin, with branches at half-integer multiples of the polarization quantum (\(\mathbf{Q}/2\)). Under inversion, \(\mathbf{P}\) maps to \(-\mathbf{P}\), which differs from \(\mathbf{P}\) by exactly one polarization quantum (\(\mathbf{Q}\)) and represents the same physical state.}
    \label{fig:polarization_symmetry}
\end{figure}

\clearpage
\newpage
\begin{figure}[htbp]
    \centering
    \includegraphics[width=1.0\textwidth]{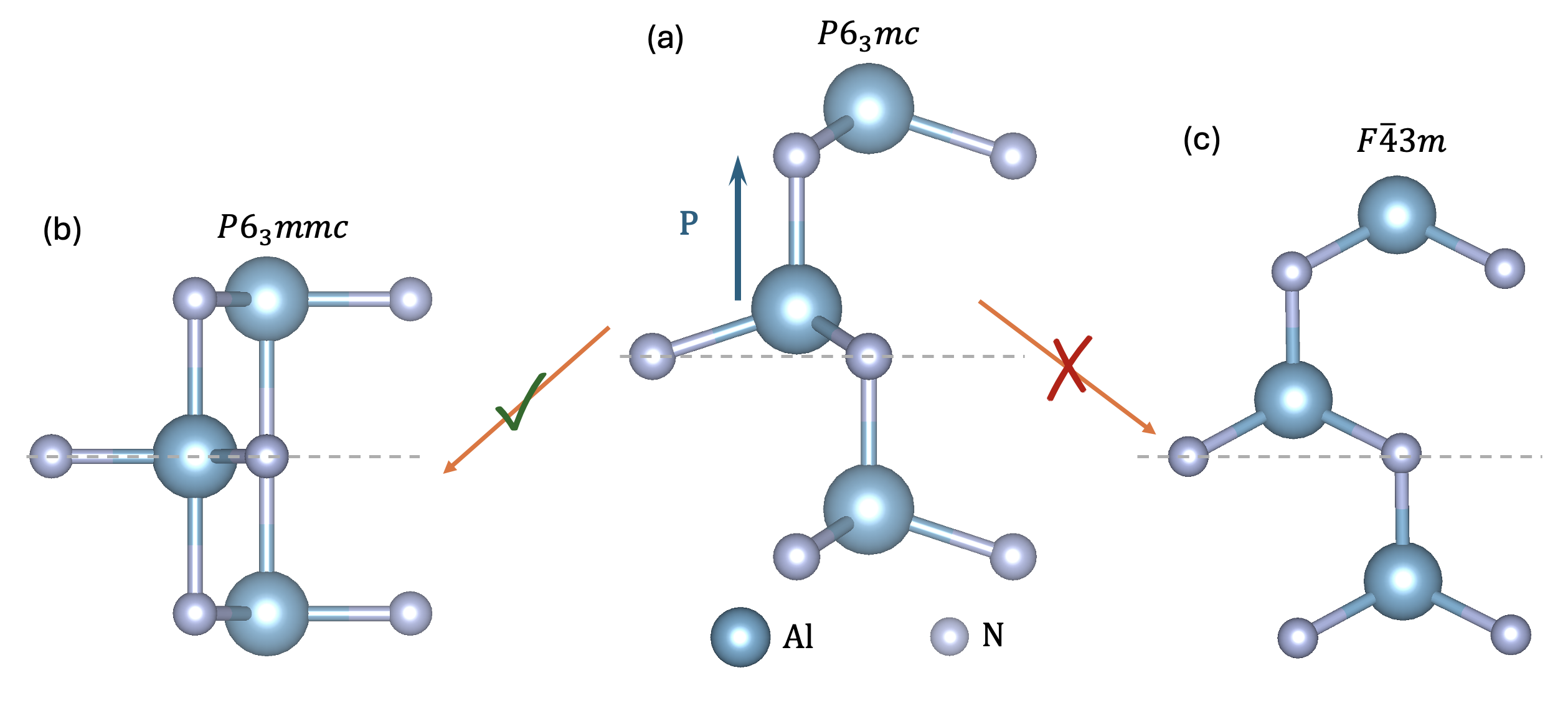}
\caption{Comparison of candidate reference structures for defining the spontaneous polarization of wurtzite AlN. (a) Polar wurtzite structure with space group \(P6_3mc\). (b) Centrosymmetric layered hexagonal structure with space group \(P6_3/mmc\). This is the appropriate reference structure because its formal polarization lattice includes zero, similar to case A in Figure 3. (c) Zinc-blende structure with space group \(F\bar{4}3m\). This structure is not an appropriate reference for a direct calculation without additional corrections because it carries a nonzero formal polarization. Moreover, it is unlikely that applying a downward electric field to structure (a) with upward polarization would drive the system into this nonpolar reference, since that would require the Al cations to move opposite to the field direction.}
    \label{fig:AlN}
\end{figure}

\clearpage
\newpage
\begin{figure}[htbp]
    \centering
    \includegraphics[width=0.5\textwidth]{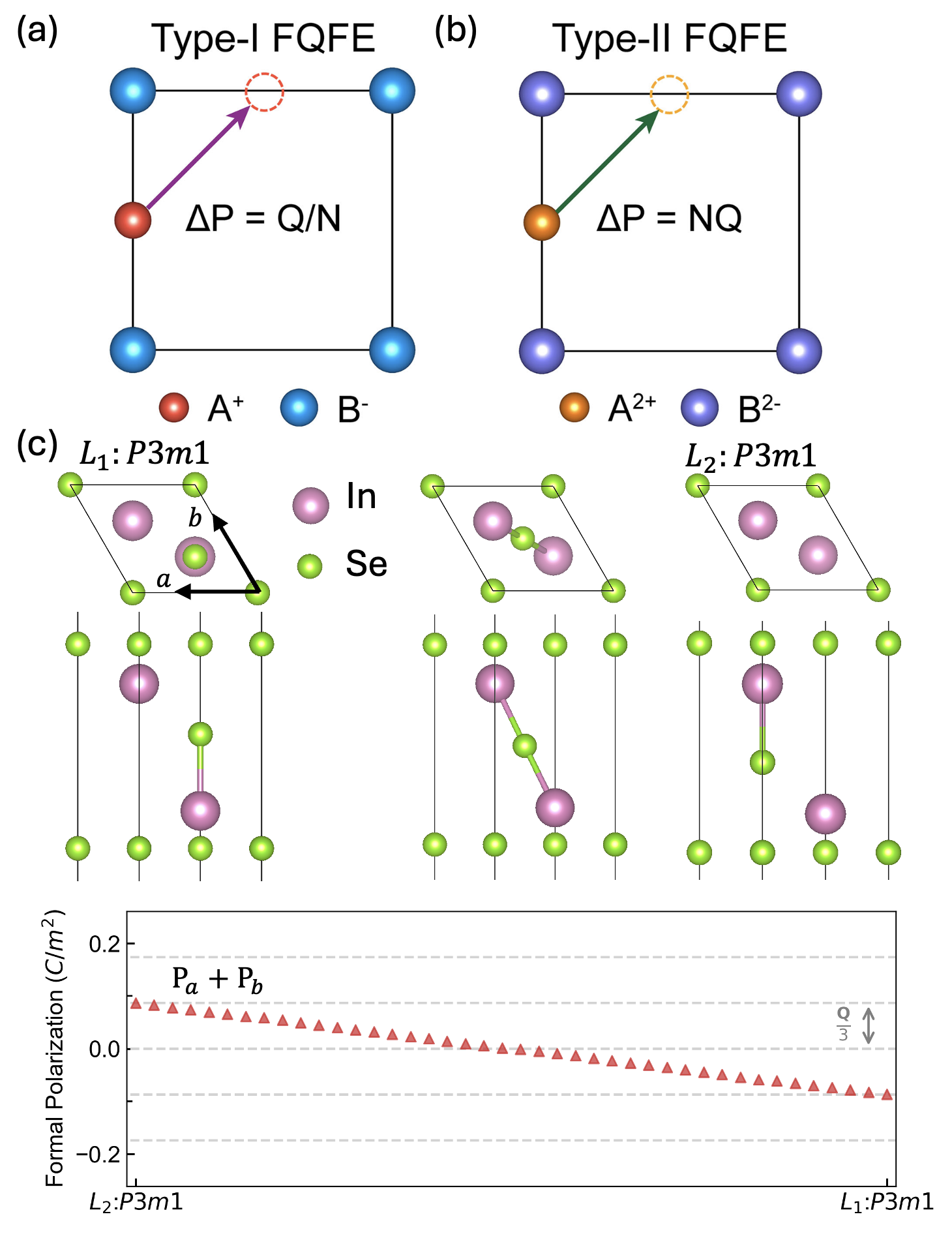}
\caption{Fractional quantum ferroelectrics and the associated polarization change. Schematics illustrating two classes of fractional quantum ferroelectrics (FQFEs). (a) Type-I FQFE: atomic displacements by a fraction of a primitive lattice vector produce a fractional polarization change, \(\Delta \mathbf{P} = \mathbf{Q}/N\), where \(\mathbf{Q} = e\mathbf{R}/\Omega\) and \(R\) is the primitive lattice vector. (b) Type-II FQFE: similar fractional displacements yield an integer multiple of the polarization quantum, \(\Delta \mathbf{P} = N\mathbf{Q}\), where \(N\) is an integer, typically due to higher ionic valence (e.g., \(A^{2+}\)) or atomic multiplicity. (c) First-principles results for the evolution of the total in-plane polarization \((\mathbf{P}_a + \mathbf{P}_b)\) in \(\alpha\)-In\(_2\)Se\(_3\) (red triangles). The top panel shows top and side views of the structures during the displacement of the middle-layer Se atom. When the Se atom is displaced by \(\mathbf{R}/3\), the polarization changes by \(-2\mathbf{Q}/3\), implying an oxidation state of \(-2\) for Se. Panel (a-b) are reproduced with permission from ref.~\cite{Pang25p116402}.}
    \label{fig:FQFE_Type}
\end{figure}

\clearpage
\newpage
\begin{figure}[htbp]
    \centering
    \includegraphics[width=1.0\textwidth]{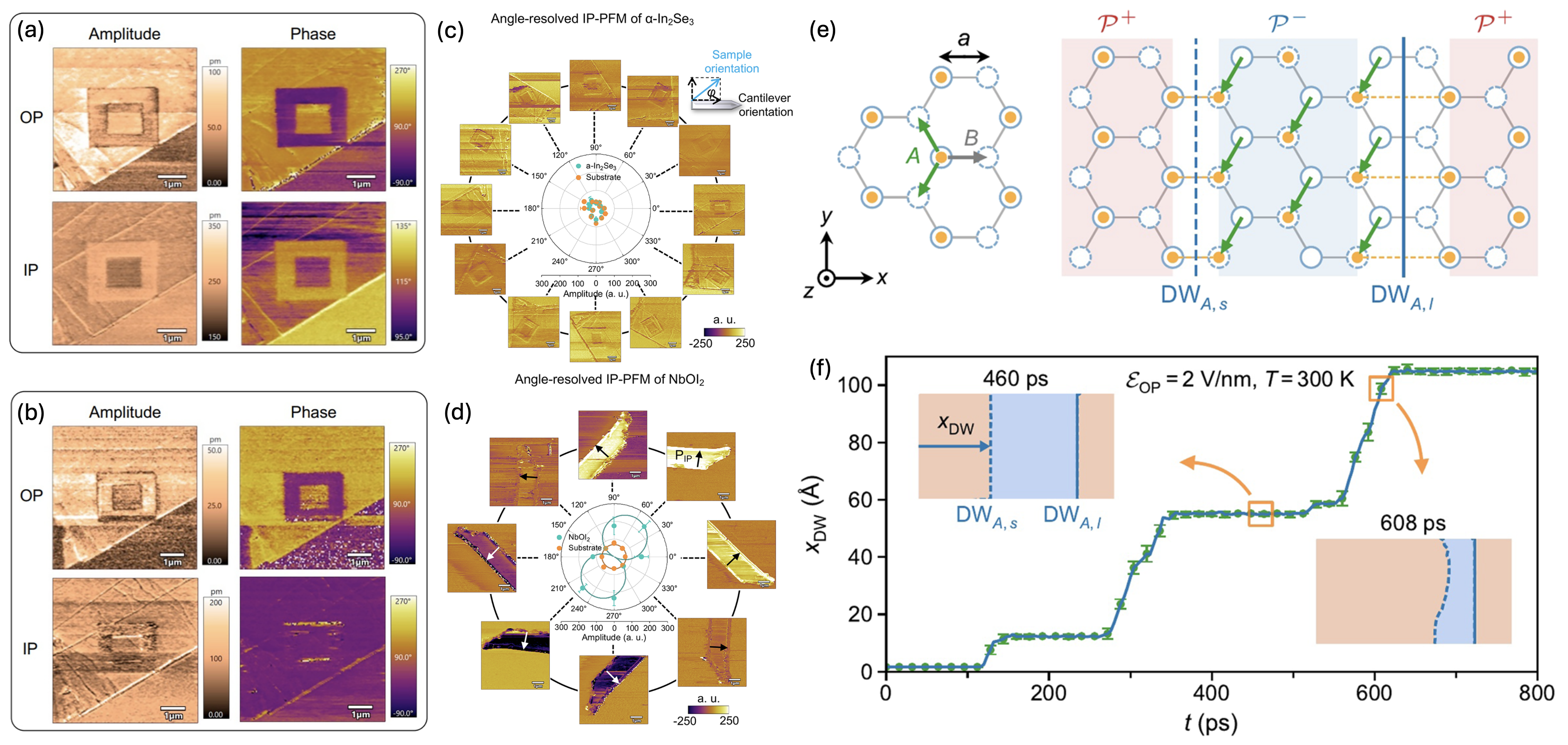}
\caption{Ferroelectricity in monolayer \(\alpha\)-In\(_2\)Se\(_3\). (a,b) Out-of-plane (OP) and in-plane (IP) PFM amplitude and phase images acquired using a stiff probe (\(k \approx 2.8~\mathrm{N/m}\)) and a soft probe (\(k \approx 0.2~\mathrm{N/m}\)), respectively. The strong reduction of the apparent IP phase contrast with the soft probe indicates that the measured IP signal mainly arises from crosstalk with the OP response. (c,d) Angle-resolved IP-PFM amplitude polar plots for \(\alpha\)-In\(_2\)Se\(_3\) and a reference NbOI\(_2\) sample. In contrast to NbOI\(_2\), \(\alpha\)-In\(_2\)Se\(_3\) shows no periodic azimuthal modulation, ruling out intrinsic in-plane piezoelectricity. (e) Atomic schematics illustrating the formation of one-dimensional domain walls in \(\alpha\)-In\(_2\)Se\(_3\), which locally break \(C_{3v}\) symmetry. (f) Temporal evolution of the one-dimensional domain-wall position, revealing abrupt and intermittent avalanche dynamics during field-driven motion under an out-of-plane electric field (\(\mathcal{E}_{\rm OP}\)). Panel (a-f) are reproduced with permission from ref.~\cite{Bai24p26103}.}
    \label{fig:InSe_FQFE}
\end{figure}

\begin{figure}[htbp]
    \centering
    \includegraphics[width=1.0\textwidth]{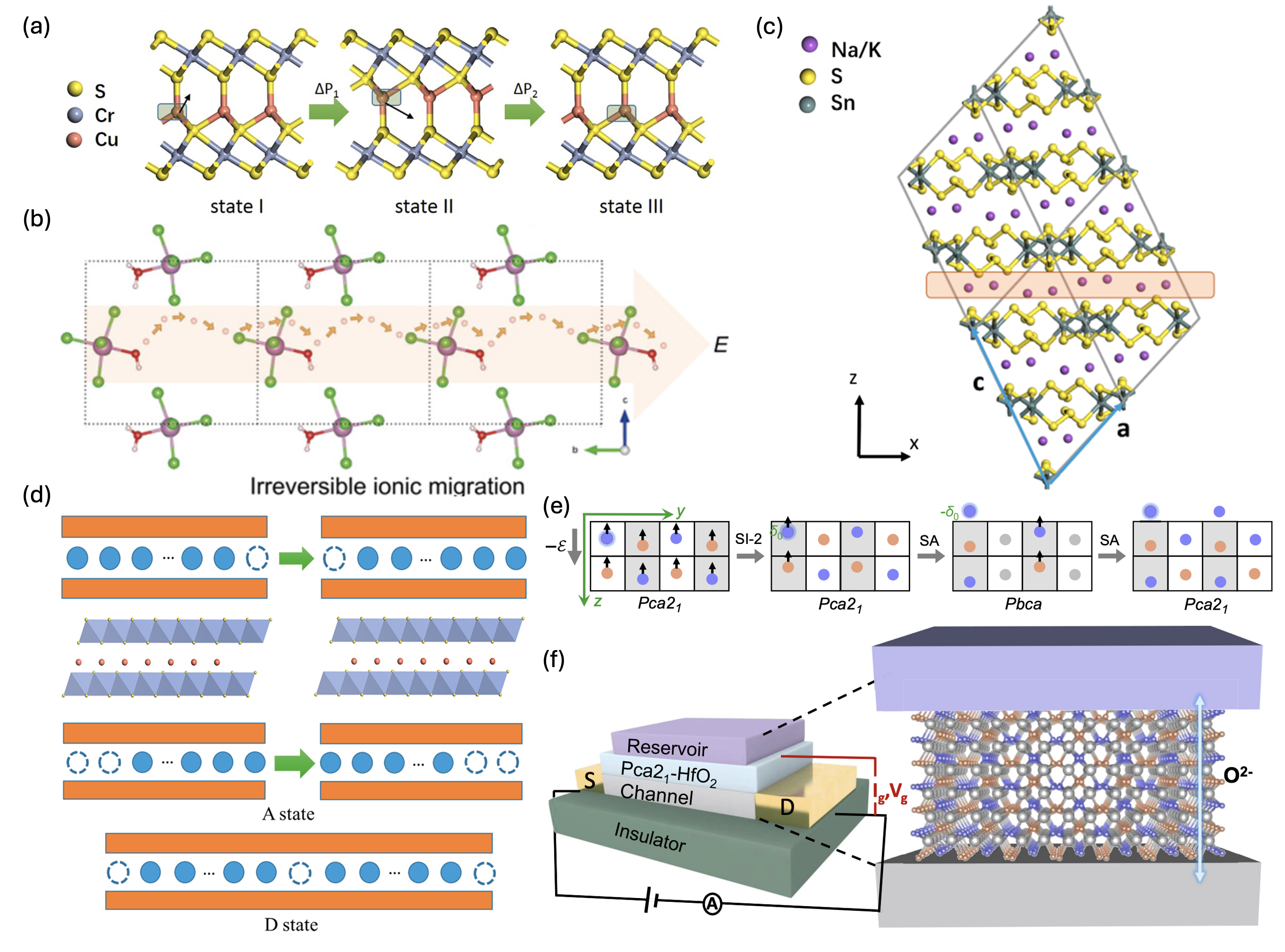}
\caption{Unconventional ferroelectricity induced by long-range ion migration. (a) Schematic illustration of the in-plane polarization changes (\(\Delta P_1\) and \(\Delta P_2\)) associated with the migration of intercalated Cu ions in CuCrS\(_2\) among different switching states (I, II, and III). (b) Irreversible proton-migration pathway along the \(b\)-axis driven by an applied electric field (\(E\)) in the organic--inorganic hybrid \((\mathrm{CETM})_2\mathrm{InCl}_5\cdot\mathrm{H_2O}\). (c) Crystal structure of the layered ionic conductor KSnS\(_4\), consisting of ion-conducting layers sandwiched between SnS\(_4\) layers. (d) Schematic models of vacancy migration from one side of an ion-conducting channel to the other. The lower panels show two possible vacancy configurations: aggregation on one side (A state) and a more uniform distribution stabilized by Coulomb repulsion (D state). (e) Polar--antipolar phase cycling in HfO\(_2\) arising from successive shift-inside (SI-2) and shift-across (SA) ferroelectric transitions under an applied electric field. (f) Schematic of an electrochemical ionic synapse device using \(Pca2_1\) HfO\(_2\) as the electrolyte layer. The enlarged view highlights ultrafast oxygen-ion (\(\mathrm{O}^{2-}\)) transport within the HfO\(_2\) layer under an electric field. Panels (a,c-d) are reproduced with permission from ref.~\cite{wang22p9552}, panel (b) is reproduced with permission from ref.~\cite{Lu23p281}, panels (e-f) are reproduced with permission from ref.~\cite{Ma23p256801}.}
    \label{fig:ICFE}
\end{figure}

\end{document}